\documentclass[a4paper]{PoS}
\usepackage{amsmath, amsthm, amssymb}

\usepackage{soul}

\newcommand{\eps}{\epsilon}
\newcommand{\taubar}{\bar\tau}
\def\rd{\mathrm{d}}

\title{Automated Calculation of Dijet Soft Functions\\ 
in the Presence of Jet Clustering Effects}

\ShortTitle{Dijet Soft Functions 
in the Presence of Jet Clustering Effects}

\author{\speaker{Guido Bell}\thanks{
Preprint numbers: SI-HEP-2018-01, QFET-2018-03, DESY 17-238.}\\
	Theoretische Physik 1, Naturwissenschaftlich-Technische Fakult\"at,
	Universit\"at Siegen, Walter-Flex-Strasse 3, 57068 Siegen, Germany\\
        E-mail: \email{bell@physik.uni-siegen.de}}

\author{Rudi Rahn\\
	Albert Einstein Center for Fundamental Physics, Institut f\"ur Theoretische Physik,
	Universit\"at Bern, Sidlerstrasse 5, 3012 Bern, Switzerland\\
        E-mail: \email{rahn@itp.unibe.ch}}
        
\author{Jim Talbert\\
        Theory Group, Deutsches Elektronen-Synchrotron (DESY), 22607 Hamburg, Germany\\
        E-mail: \email{james.talbert@desy.de}}       

\abstract{We extend our framework for the automated calculation of dijet soft functions
to observables that do not obey the non-Abelian exponentiation theorem,  
like jet-veto or grooming soft functions that are sensitive to clustering effects of 
the jet algorithm. Although the matrix element for uncorrelated double emissions has a 
simpler structure than the one for correlated emissions, we argue that its singularity 
structure poses more stringent constraints on the required phase-space parametrisation. 
Our algorithm applies to both SCET-1 and SCET-2 soft functions and it is implemented 
in the novel program {\tt SoftSERVE}. We present results for various jet-veto observables 
and obtain new predictions for the soft-drop jet-grooming algorithm.}

\FullConference{13th International Symposium on Radiative Corrections\\
		24-29 September, 2017\\
		St. Gilgen, Austria}

\begin{document}

\section{Introduction}

In this contribution we report on progress in our effort to automate the calculation 
of dijet soft functions to next-to-next-to-leading order (NNLO) accuracy. The general idea 
of our framework was introduced in~\cite{Bell:2015lsf}, to which we refer for most of the 
definitions that are used in this article. The extensions with respect 
to~\cite{Bell:2015lsf} are threefold: 
(i) We relax the assumption that the observable should be consistent with the 
non-Abelian exponentiation (NAE) theorem.
(ii) The formalism is extended to soft functions that are relevant for 
transverse-momentum resummation (so-called SCET-2 observables). 
(iii) We introduce the program {\tt SoftSERVE}, which allows for an efficient numerical 
evaluation of soft functions that are defined in terms of two back-to-back light-like 
Wilson lines. We briefly address each of these points in the following, 
before presenting sample results for jet-veto and jet-grooming soft functions.

\section{Uncorrelated emissions}
\label{sec:CF2}

The soft functions we consider are of the form
\begin{equation}
S(\tau, \mu) = \frac{1}{N_c} \; \sum_X \; 
\mathcal{M}(\tau;\lbrace k_{i} \rbrace)\;
\mathrm{Tr}\; 
\langle 0 | S^{\dagger}_{\bar{n}} S_{n} | X \rangle
\langle X | S^{\dagger}_{n} S_{\bar{n}} | 0 \rangle \,,
\end{equation}
where $S_{n}$ and $S_{\bar{n}}$ are soft Wilson lines in the fundamental colour 
representation. Up to NNLO it is irrelevant whether the light-like vectors $n^\mu$ and 
$\bar n^\mu$ (which are normalised to $n\cdot \bar n=2$) correspond to incoming or 
outgoing directions~\cite{Catani:2000pi,Kang:2015moa}. Our results therefore equally 
apply to $e^+e^-$ dijet observables, one-jet observables in deep-inelastic scattering 
or zero-jet observables at hadron colliders. For simplicity, we refer to all of these 
cases as \emph{dijet soft functions} in the following.

The definition involves a generic measurement function 
$\mathcal{M}(\tau,\lbrace k_{i} \rbrace)$ that provides a constraint on the soft 
radiation with parton momenta $\lbrace k_{i} \rbrace$ according to the observable 
under consideration. The explicit form we assume for the single-emission measurement
function was given in~\cite{Bell:2015lsf}. It depends on the Laplace variable $\tau$ 
(of dimension 1/mass) and a function $f(y,\theta)$ that encodes the angular and
rapidity dependence of the observable. In our approach, it is crucial that this function
is finite and non-zero in the limits in which the corresponding matrix element becomes
singular. We therefore factor out an appropriate power of the rapidity variable
$y$ that is controlled by a parameter $n$ (see~\cite{Bell:2015lsf} for details). 

At NNLO the double real-emission contribution consists of three colour structures:
$C_{F}C_{A}$, $C_{F}T_{F}n_{f}$ and $C_{F}^{2}$. The parametrisation of the measurement 
function for the first two structures (the so-called correlated-emission contribution) 
was specified in~\cite{Bell:2015lsf}. For the remaining colour structure, we start from
\begin{equation}
S_2^{RR}(\eps)  = \frac{(4\pi e^{\gamma_E} \tau^2)^{-2\eps}}{(2\pi)^{2d-2}} \,
\int d^{d}k \;\, \delta(k^{2}) \,\theta(k^{0}) \,
\int d^{d}l \;\, \delta(l^{2}) \,\theta(l^{0}) 
\;\mathcal{M}(\tau; k,l) \, |\mathcal{A}_{RR}(k,l)|^{2} 
\end{equation}
in dimensional regularisation with $d=4-2\eps$. The matrix element of the 
uncorrelated-emission contribution is particularly simple,
\begin{equation}
|\mathcal{A}_{RR}(k,l)|^{2} =  
\frac{2048 \pi^{4} \,C_{F}^2}{k_+ k_- l_+ l_-} \,.
\end{equation}
In~\cite{Bell:2015lsf} we assumed that the measurement function was consistent with 
the NAE theorem~\cite{Gatheral:1983cz,Frenkel:1984pz} and the 
uncorrelated-emission contribution was therefore proportional to the square of the NLO correction. 
We relax this assumption in this work and compute this contribution explicitly. 
To this end, we need to find a phase-space parametrisation that disentangles the 
singularity structure of the matrix element and that allows us to control the measurement 
function in its singular limits (in the same sense that the NLO function $f(y,\theta)$ 
has to be finite and non-zero as $y\to 0$). It turns out that the latter requirement 
necessarily introduces non-trivial correlations between the parton momenta $k$ and $l$. 
Our parametrisation for the uncorrelated-emission contribution,
\begin{align}
q_T &= \sqrt{k_+ k_-} 
\left( \frac{\sqrt{l_+ l_-}}{l_-+l_+} \right)^{-n}
+ \sqrt{l_+ l_-} 
\left( \frac{\sqrt{k_+ k_-}}{k_-+k_+} \right)^{-n} ,\hspace{1.0cm}
y_k = \frac{k_{+}}{k_{-}} \,,\qquad
\nonumber\\
b &= \sqrt{\frac{k_{+}k_{-}}{l_{+}l_{-}}} 
\left( \frac{\sqrt{k_+ k_-}}{k_-+k_+} \right)^{n}
\left( \frac{\sqrt{l_+ l_-}}{l_-+l_+} \right)^{-n}
 , \hspace{2.7cm}
y_l =\frac{l_{+}}{l_{-}} \,,
\end{align}
is therefore more complicated than the one we used for the correlated emissions
(see~\cite{Bell:2015lsf}). Notice that this parametrisation depends on the parameter $n$, 
i.e.~it is strictly speaking not observable-independent. In physical terms, the 
variables $y_k$ and $y_l$ are measures of the rapidities of the individual partons, 
whereas $b$ and $q_T$ reduce for $n=0$ to the ratio and the scalar sum of their 
transverse momenta, respectively (the parentheses introduce rapidity-dependent
weight factors for $n\neq0$). The measurement function for the uncorrelated 
emission contribution is then parametrised as
\begin{align}
\mathcal{M}(\tau; k,l) = \exp\big(-\tau\, q_{T}\, y_k^{n/2}\, y_l^{n/2}\, 
G(y_k,y_l,b,\theta_k,\theta_l,\theta_{kl})\big)\,,
\end{align}
where $\theta_k$, $\theta_l$ and $\theta_{kl}$ are angular variables that were 
defined in~\cite{Bell:2015lsf}. The linear dependence on $q_T$ is fixed on 
dimensional grounds, and the factors $y_k^{n/2}$ and $y_l^{n/2}$ are required to
make the function $G(y_k,y_l,b,\theta_k,\theta_l,\theta_{kl})$ finite and non-zero
in the collinear limits $y_k\to0$ and $y_l\to0$. As an example, we quote the
measurement function for $W$-production at large transverse momentum~\cite{Becher:2012za},
\begin{align}
G(y_k,y_l,b,\theta_k,\theta_l,\theta_{kl})=
\frac{b(1+y_l)}{(1+b)} \,\big( 1 + y_k - 2 \sqrt{y_k}\, \cos\theta_k \big)
+\frac{(1+y_k)}{(1+b)} \,\big( 1 + y_l - 2 \sqrt{y_l}\, \cos\theta_l \big)\,.
\end{align}
Similar to the correlated-emission contribution, the measurement function 
satisfies non-trivial constraints from infrared-collinear safety,
\begin{align}
G(y_k,y_l,0,\theta_k,\theta_l,\theta_{kl}) = 
\frac{f(y_l,\theta_l)}{(1+y_k)^n}\,,\qquad\qquad
G(y_l,y_l,b,\theta_l,\theta_l,0) = \frac{f(y_l,\theta_l)}{(1+y_l)^n}\,,
\end{align}
which correspond to the soft limit $k^\mu\to0$ and the collinear limit 
$k^\mu\propto l^\mu$, respectively. We further exploit the symmetries from 
$n\leftrightarrow\bar n$ and $k\leftrightarrow l$ exchange to map the integration 
region onto the unit hypercube. For the jet algorithms we have considered so far, 
it is also crucial to disentangle the scalings of the measurement function
in the joint limit $y_k\to0$ and $y_l\to0$ at a fixed ratio $r=y_k/y_l$ from the one 
of the subsequent limits with $r\to0$. In total we are then left with a six-dimensional 
integral representation of the uncorrelated-emission contribution, which contains an 
explicit singularity from the limit $q_T\to0$ and implicit divergences from $y_k\to0$, 
$r\to0$ and $b\to0$. The $C_{F}^2$ contribution thus starts with a $1/\eps^4$ pole. 

\section{SCET-2 and collinear anomaly}

The algorithm we have outlined so far leads to rapidity integrals of the form 
$\int_0^1 dy \;\, y^{-1+n\eps}$, which are not regularised for $n=0$. This 
particular case corresponds to a SCET-2 observable, which are known to require an 
additional regulator on top of dimensional regularisation. Here we follow the 
strategy proposed in~\cite{Becher:2011dz} and implement the regulator on 
the level of the phase-space integrals. In order to keep the $n\leftrightarrow\bar n$ 
symmetry, we write the generic phase-space measure as
\begin{equation}
\int\!d^dp \; \left(\frac{\nu}{p_++p_-}\right)^\alpha \;  \delta(p^2) \theta(p^0) \,,
\end{equation}
and the rapidity divergences then manifest themselves as poles in the regulator
$\alpha$.

The $1/\alpha$ poles induce logarithmic corrections in the rapidity scale $\nu$, 
which are controlled by the collinear anomaly exponent $F$,
\begin{align}
S(\tau,\mu,\nu)
&= (\nu^2\taubar^2)^{-F(\tau,\mu)} \;W_S(\tau,\mu)\,,
\end{align}
where $\taubar = \tau e^{\gamma_E}$ and the soft remainder function $W_S$ contains 
the finite terms in the $\alpha$-expansion. From the calculation of the bare soft function,
one can thus determine the bare collinear 
anomaly exponent $F_0$, which renormalises additively in Laplace space, $F_0 = F + Z_F$. 
The renormalised anomaly 
exponent $F$ fulfills a renormalisation group (RG) equation 
\begin{align}
\frac{\rd}{\rd \ln\mu}  \; F(\tau,\mu)
&= 2 \,\Gamma_{\mathrm{cusp}}(\alpha_s)\,,
\end{align}
which is governed by the cusp anomalous dimension. Expanding 
$\Gamma_{\mathrm{cusp}}(\alpha_s)= \sum_{n=0}^\infty \,\Gamma_n (\frac{\alpha_s}{4\pi})^{n+1}$, 
the two-loop solution of the RG equation takes the form
\begin{align}
F(\tau,\mu) &= 
\left( \frac{\alpha_s}{4 \pi} \right) 
\Big\{ 2\Gamma_0 \,L 
+ d_1 \Big\}
+\left( \frac{\alpha_s}{4 \pi} \right)^2 
\Big\{ 2 \beta_0\Gamma_0\, L^2 
 + 2 \left( \Gamma_1 + \beta_0 d_1 \right) L + d_2 \Big\}
 \label{eq:d1d2}
\end{align}
with $L=\ln(\mu\taubar)$. Explicit expressions of the expansion coefficients $\Gamma_0$ 
and $\Gamma_1$ as well as the beta-function coefficient $\beta_0$ can be found 
in~\cite{Bell:2015lsf}. 

The $Z$-factor $Z_{F}$ satisfies a similar RG equation as the anomaly exponent
and its explicit form to two-loop order is given by
\begin{align} 
Z_{F} &= \left( \frac{\alpha_s}{4 \pi} \right) 
\left\{ \frac{\Gamma_0}{\eps} \right\}
+\left( \frac{\alpha_s}{4 \pi} \right)^2 
\bigg\{ - \frac{\beta_0\Gamma_0}{2\eps^2}
+ \frac{\Gamma_1}{2\eps}
\bigg\}\,.
\end{align}
The cancellation of the divergences $1/\eps^j$ with $j=1,2$ in the renormalised 
anomaly exponent $F$ provides a strong check of our calculation. The finite terms,
on the other hand, determine the one-loop and two-loop anomaly coefficients $d_1$
and $d_2$. The extraction of $d_2$ from the bare soft function is actually subtle 
since the one-particle and two-particle cuts have different scalings in the rapidity 
scale $\nu$. The coefficients of the $1/\alpha$ pole terms and the 
associated logarithms in the rapidity scale $\nu$ are therefore different; 
see the discussion in section 4.3 of~\cite{Becher:2012qc} for more details.
 
\section{Numerical implementation}

The integral representations that we have derived above and in~\cite{Bell:2015lsf} 
are implemented in the custom {\tt C++} program {\tt SoftSERVE}. In~\cite{Bell:2015lsf} 
we found an overlapping divergence in the correlated-emission contribution 
(in the limits $a\to 1$ and $t_{kl}\to 0$), which we could resolve in the meantime 
by an additional substitution. In other words, all singularities are factorised in 
the applied formulae and no sector decomposition strategy is needed anymore to 
disentangle the divergences. 

With all singularities factorised in the form $x^{-1+m\eps+k\alpha}$, they can easily
be made explicit by introducing standard plus-distributions. One must further respect
the ordering of the $\alpha$- and $\eps$-expansions, since the additional regulator should
only be used to regularize rapidity divergences for SCET-2 soft functions. As the 
measurement functions $f$, $F$ and $G$ are by construction finite and non-zero in the 
singular limits (and independent of the regulators), they can be kept symbolic during 
the subtraction and expansion steps, and their explicit forms are only resolved at the 
final numerical integration stage.

For the numerical integrations {\tt SoftSERVE} applies the Divonne integrator of the 
Cuba library~\cite{Hahn:2004fe}. The code contains a number of further refinements 
to improve the convergence of the numerical integrations (for more details, 
see~\cite{Rahn:thesis}). In particular, we eliminated all (integrable) square-root 
divergences with appropriate remappings in order to obtain a more reliable error estimate 
for the Monte Carlo integrations. For some of the observables, we observed large 
cancellations between different terms, which could not be resolved with double 
precision variables. We therefore implemented an option to work with multi-precision 
variables provided by the {\tt boost}~\cite{boost} and {\tt GMP/MPFR}~\cite{GMP/MPFR} 
libraries, although this option significantly slows down the program.

Leaving the last issue aside, a typical {\tt SoftSERVE} run usually takes less than 
half an hour on a standard quad-core machine to determine a bare NNLO soft
function up to the finite terms to 3-4 digits precision. As the Cuba integrators 
support parallelisation, it is possible to increase the precision to a 
few more digits on a reasonable time scale. In addition, {\tt SoftSERVE} provides 
scripts for the renormalisation of  bare soft functions in both Laplace and cumulant 
space; further details will be given in a future publication~\cite{BRT}.

We checked our results with an independent code that uses the public
programs {\tt SecDec~\!\!\!3}~\cite{Borowka:2015mxa}  and {\tt pySecDec}~\cite{Borowka:2017idc}.
As the new python-based version of {\tt SecDec} supports an arbitrary number of 
analytic regulators, it is  particularly suited for SCET-2 problems. 
For the numerical integrations in {\tt SecDec} we used the Cuba integrators
{\tt Suave} and {\tt Vegas} for independent cross checks.

\begin{figure}[t]
\includegraphics[scale=0.255]{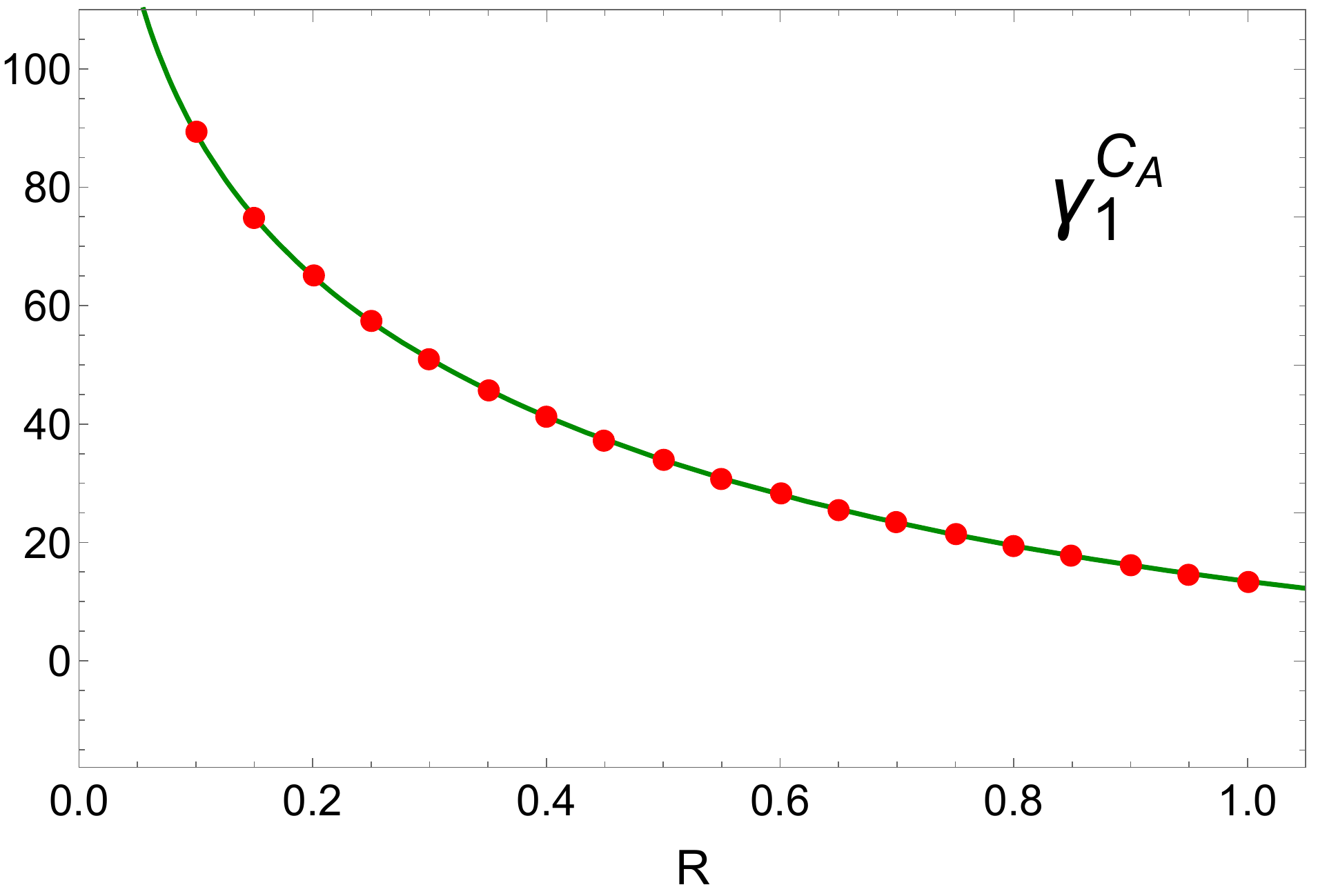}
\hspace{0.5mm}
\includegraphics[scale=0.249]{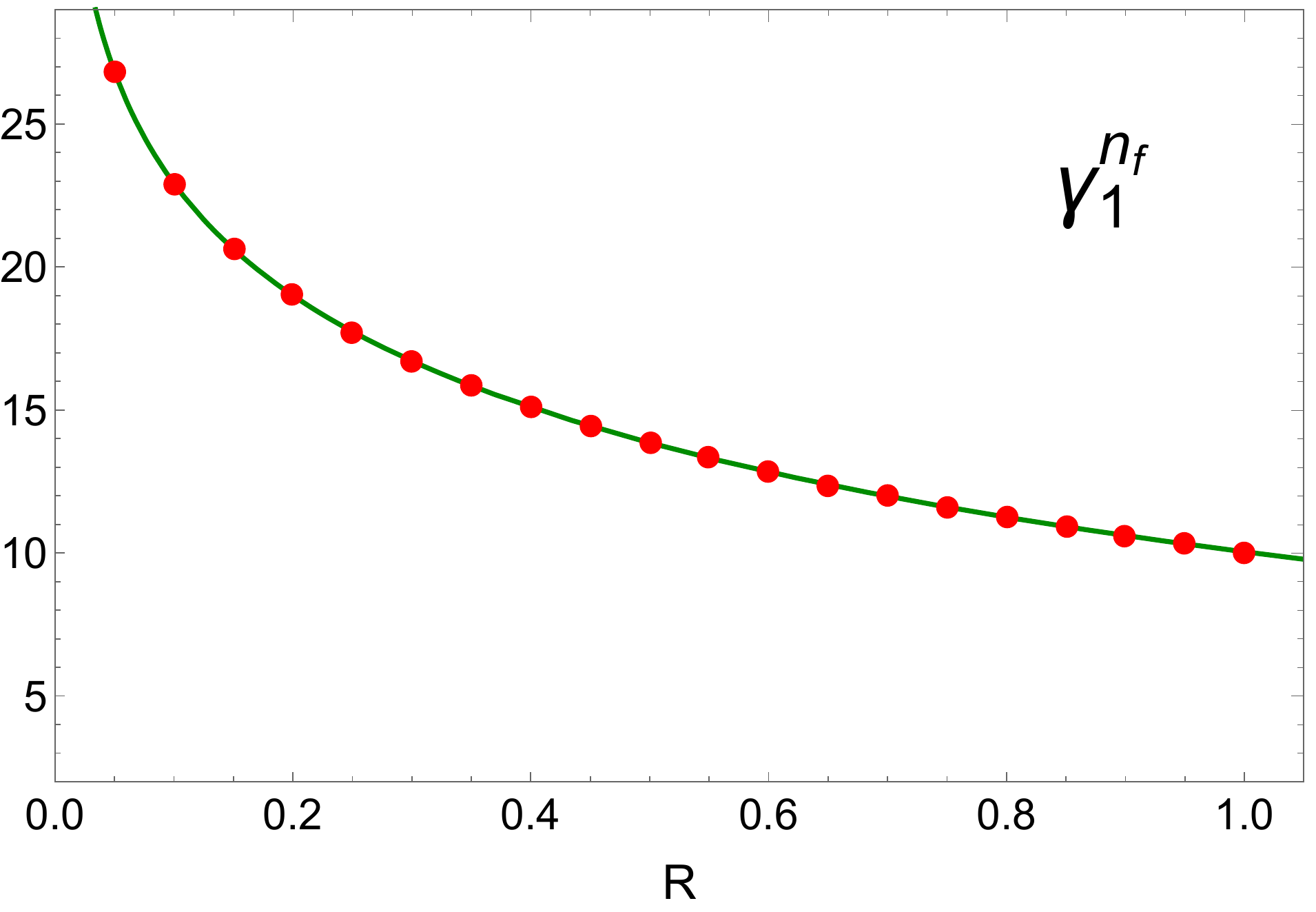}
\hspace{0.5mm}
\includegraphics[scale=0.255]{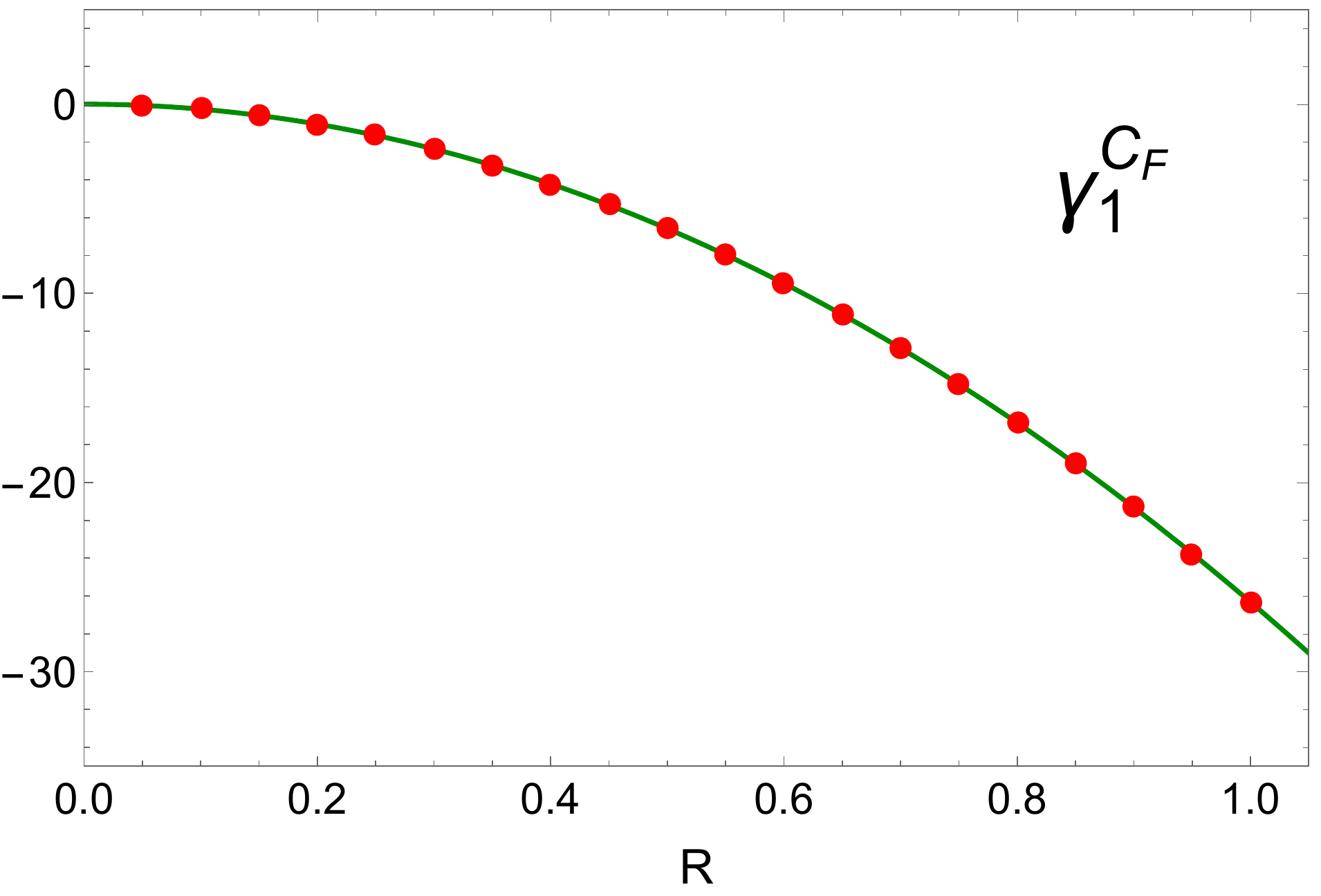}
\caption{Two-loop anomalous dimension of the rapidity-dependent jet-veto 
soft functions from~\cite{Gangal:2014qda}. The (red) dots represent our {\tt SoftSERVE} 
numbers and the solid (green) lines are the fit functions that were obtained 
in~\cite{Gangal:2016kuo}.}
\label{fig:jetveto}
\end{figure}

\section{Results}

In this article we focus on jet-veto and jet-grooming observables. For SCET-1 soft functions with $n\neq0$, we present
results for the two-loop  anomalous dimension $\gamma_1^S$ that was defined 
in~\cite{Bell:2015lsf}. For observables that violate the NAE theorem, the anomalous
dimension has three colour structures,
\begin{align} 
\gamma_1^S &= \gamma_1^{C_A} \;C_F C_A 
+ \gamma_1^{n_f} \;C_F T_F n_f
+ \gamma_1^{C_F} \;C_F^2\,,
\end{align}
of which $\gamma_1^{C_A}$ and $\gamma_1^{n_f}$ can be calculated with the strategy 
from~\cite{Bell:2015lsf}. The NAE-violating coefficient $\gamma_1^{C_F}$,
on the other hand, can be determined with the novel method that we discussed
in Section~\ref{sec:CF2}.

We first consider the rapidity-dependent jet-veto observables that were introduced 
in~\cite{Gangal:2014qda}\footnote{Notice that the four jet-veto observables that were 
considered in~\cite{Gangal:2014qda} have the same soft anomalous dimension.}. 
As the RG equation for these observables holds in cumulant rather than Laplace
space, we have to take the Laplace transform of the respective soft 
functions to bring them into the form that we assume for the measurement function.
It is then possible to correct for the factors associated with the inversion of the
Laplace transformation on the level of the bare soft functions.

For the one-loop soft anomalous dimension, the integrals can be solved analytically
and one finds $\gamma_0^{S}=0$. At two-loop order, we have evaluated the soft functions
with {\tt SoftSERVE} for 20 values of the jet radius $R$ between $R=0.05$ and $R=1$.
Our results are displayed in Figure~\ref{fig:jetveto}, which also shows fitting
functions that were obtained in a previous calculation~\cite{Gangal:2016kuo}. Our results 
nicely agree with these functions and they represent the first confirmation 
of the calculation in~\cite{Gangal:2016kuo}\footnote{The uncertainties
of our numerical predictions are too small to be visible in the plots that we show in this article.}.

\begin{figure}[t]
\includegraphics[scale=0.253]{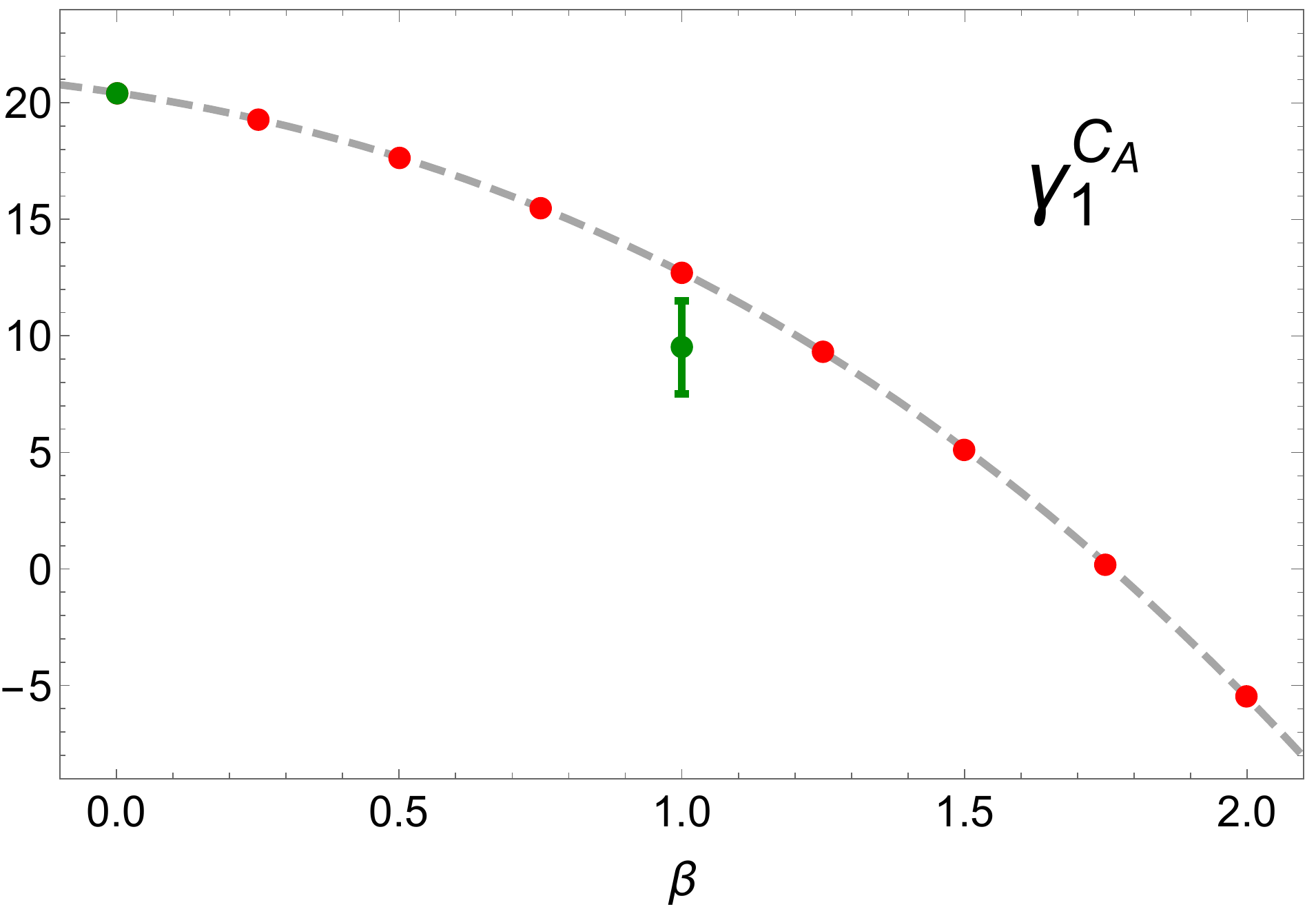}
\hspace{0.5mm}
\includegraphics[scale=0.251]{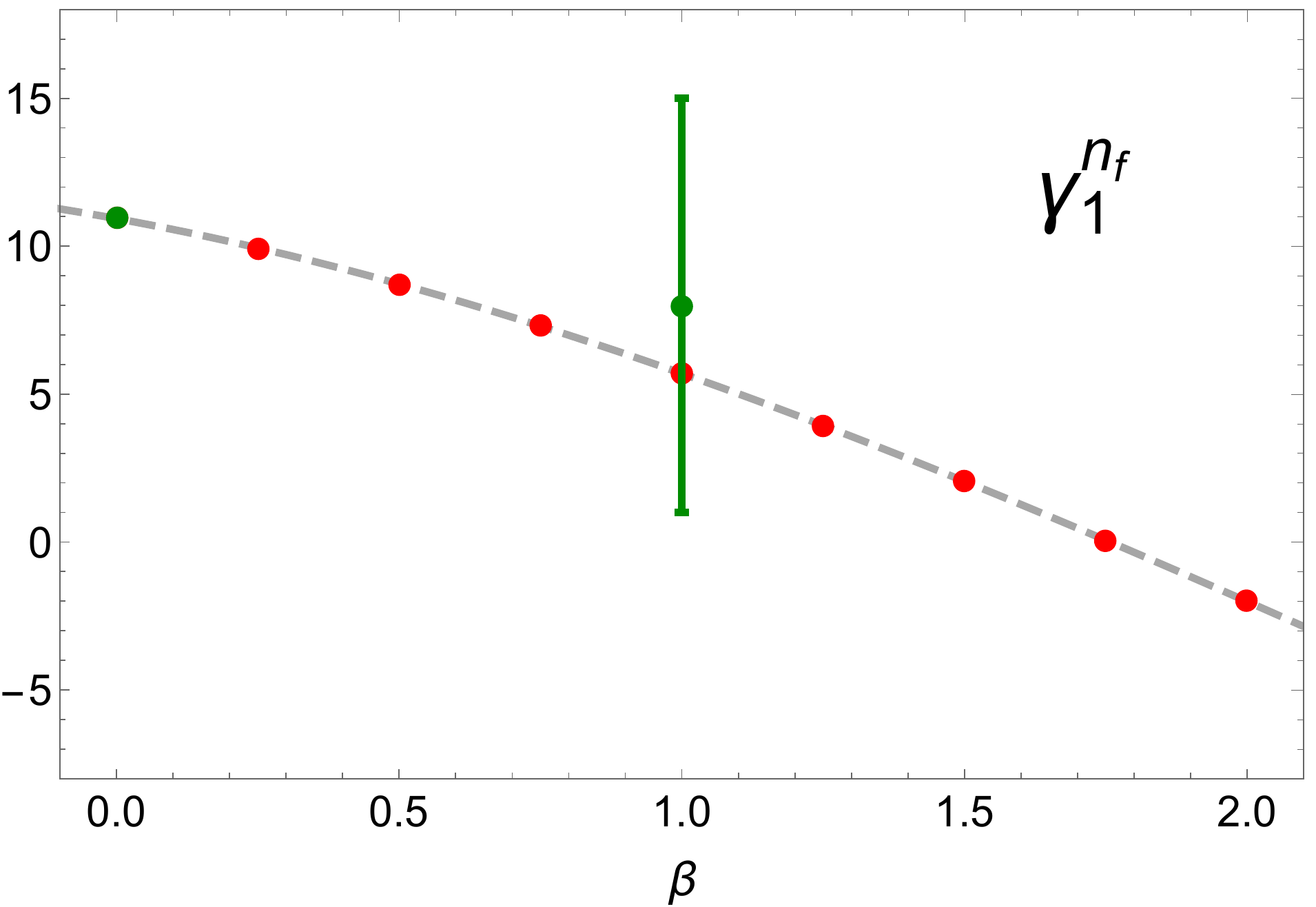}
\hspace{0.5mm}
\includegraphics[scale=0.253]{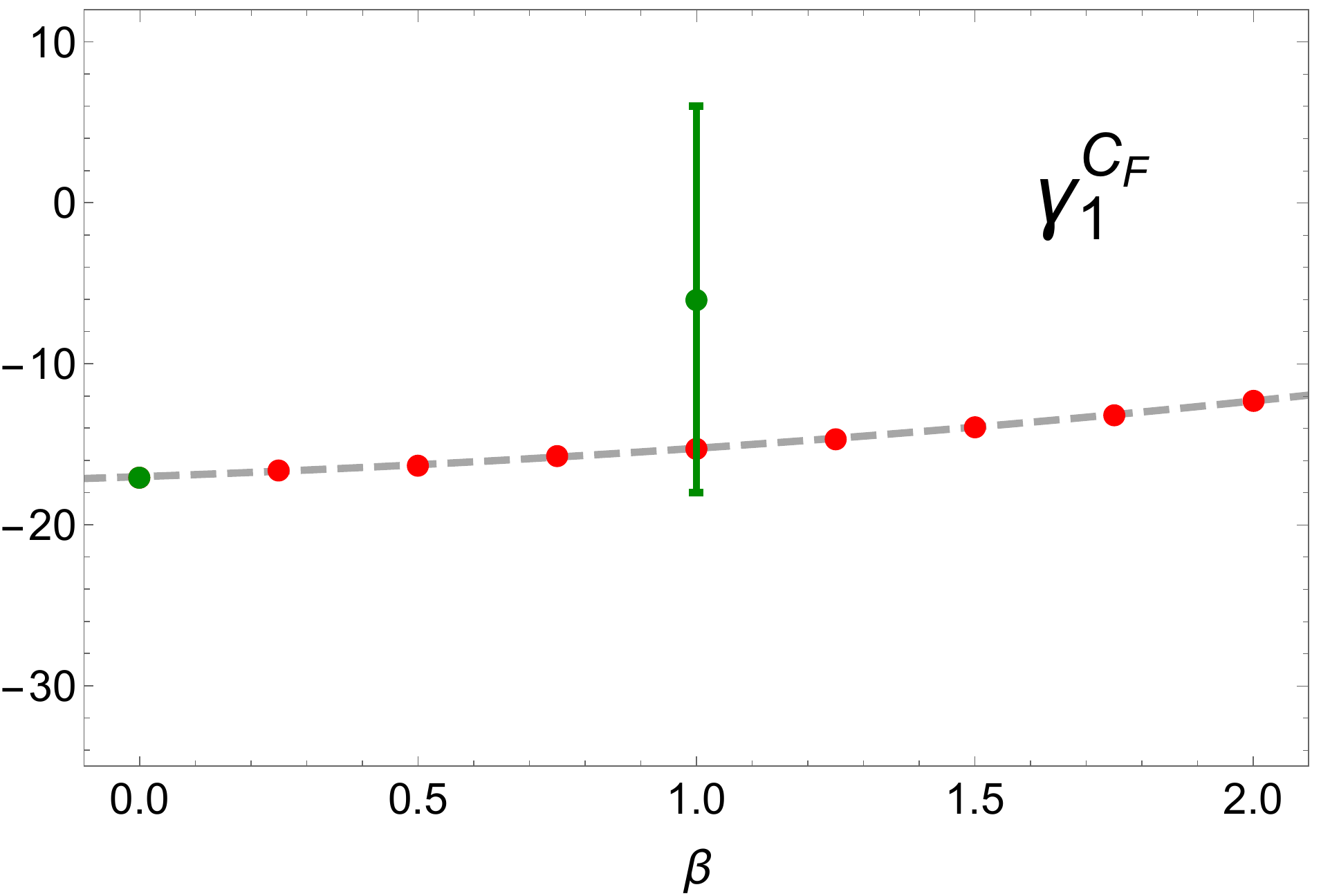}
\caption{Two-loop anomalous dimension of the soft-drop jet-grooming
soft function from~\cite{Frye:2016aiz}. The (red) dots represent our {\tt SoftSERVE} 
numbers and the dashed lines are interpolating functions of our results. 
The (green) dots with error bars are the numbers from~\cite{Frye:2016aiz}.}
\label{fig:jetgrooming}
\end{figure}

We next turn to the soft-drop jet-grooming soft function that was discussed 
in~\cite{Frye:2016aiz}. This function also renormalises multiplicatively in 
cumulant space and one again finds $\gamma_0^{S}=0$ at NLO. At NNLO
we have evaluated the soft function with {\tt SoftSERVE} for nine values 
of the parameter $\beta$, which controls the aggressiveness of the jet groomer.
Our results are shown in Figure~\ref{fig:jetgrooming} together with the numbers 
from~\cite{Frye:2016aiz}. In this work the authors extracted the anomalous 
dimension from an analytic calculation for $\beta=0$ and our results again
nicely confirm these numbers. For $\beta=1$, on the other hand, the authors 
extracted $\gamma_1^{S}$ from a fit to the {\tt EVENT2} generator.
As is evident from the plots, our results agree with these numbers
but they are far more precise. In addition, the computing time for 
running {\tt SoftSERVE} is several orders of magnitude smaller than the one that 
was used for the {\tt EVENT2} fits. We can therefore compute the anomalous
dimension for various values of $\beta$ to obtain interpolating functions
that are also shown in the figure. 

\begin{figure}[t]
\includegraphics[scale=0.253]{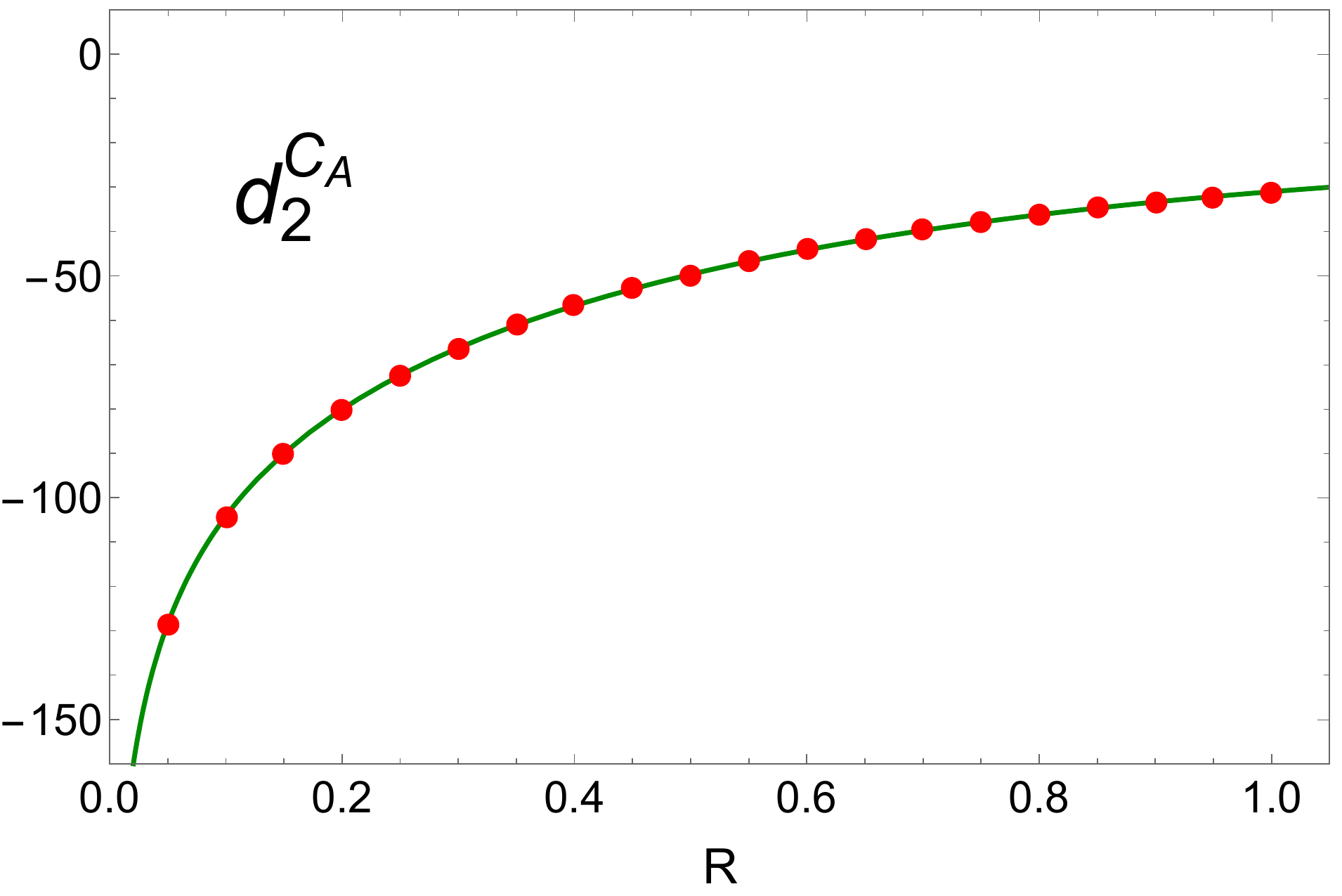}
\hspace{0.5mm}
\includegraphics[scale=0.251]{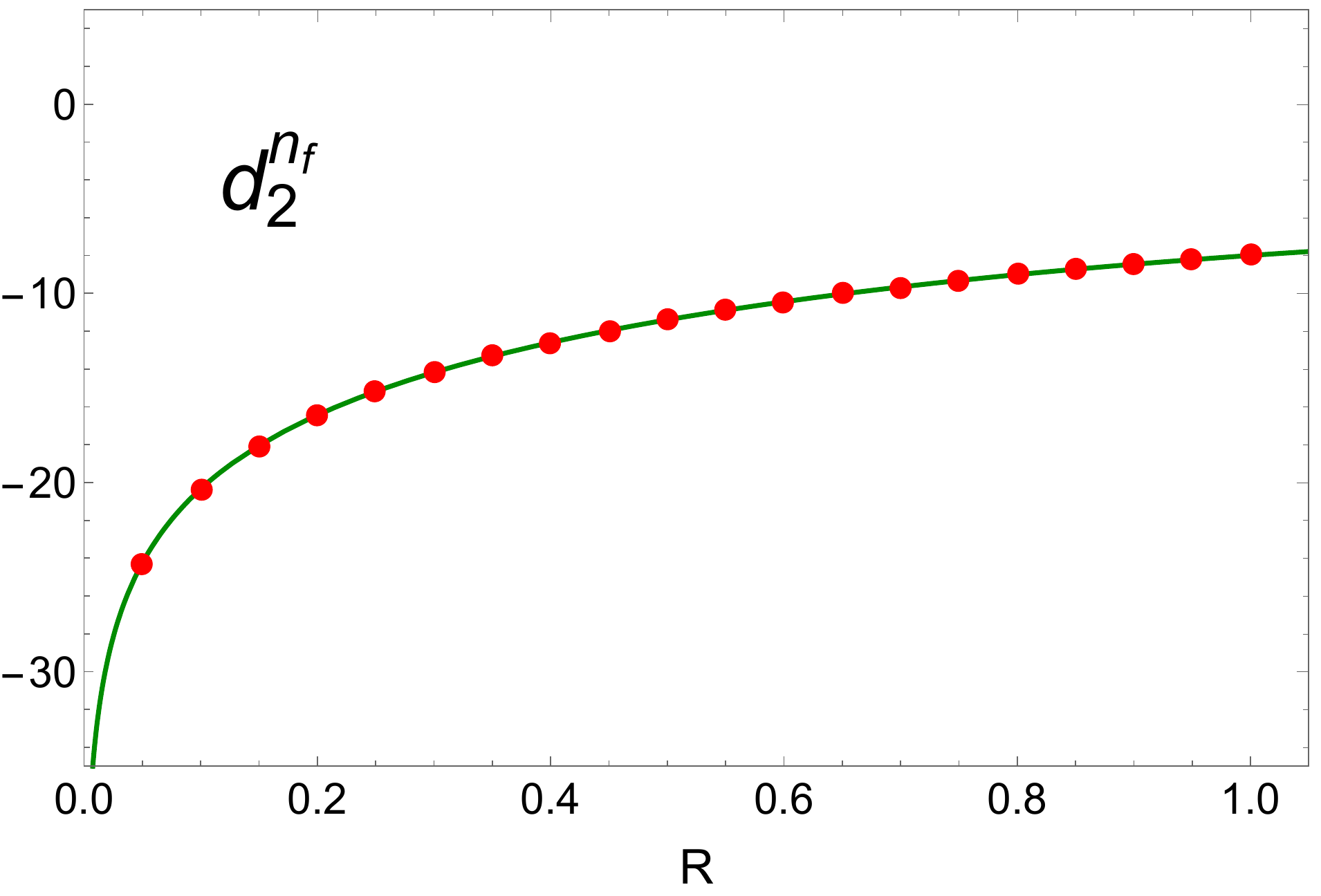}
\hspace{0.5mm}
\includegraphics[scale=0.253]{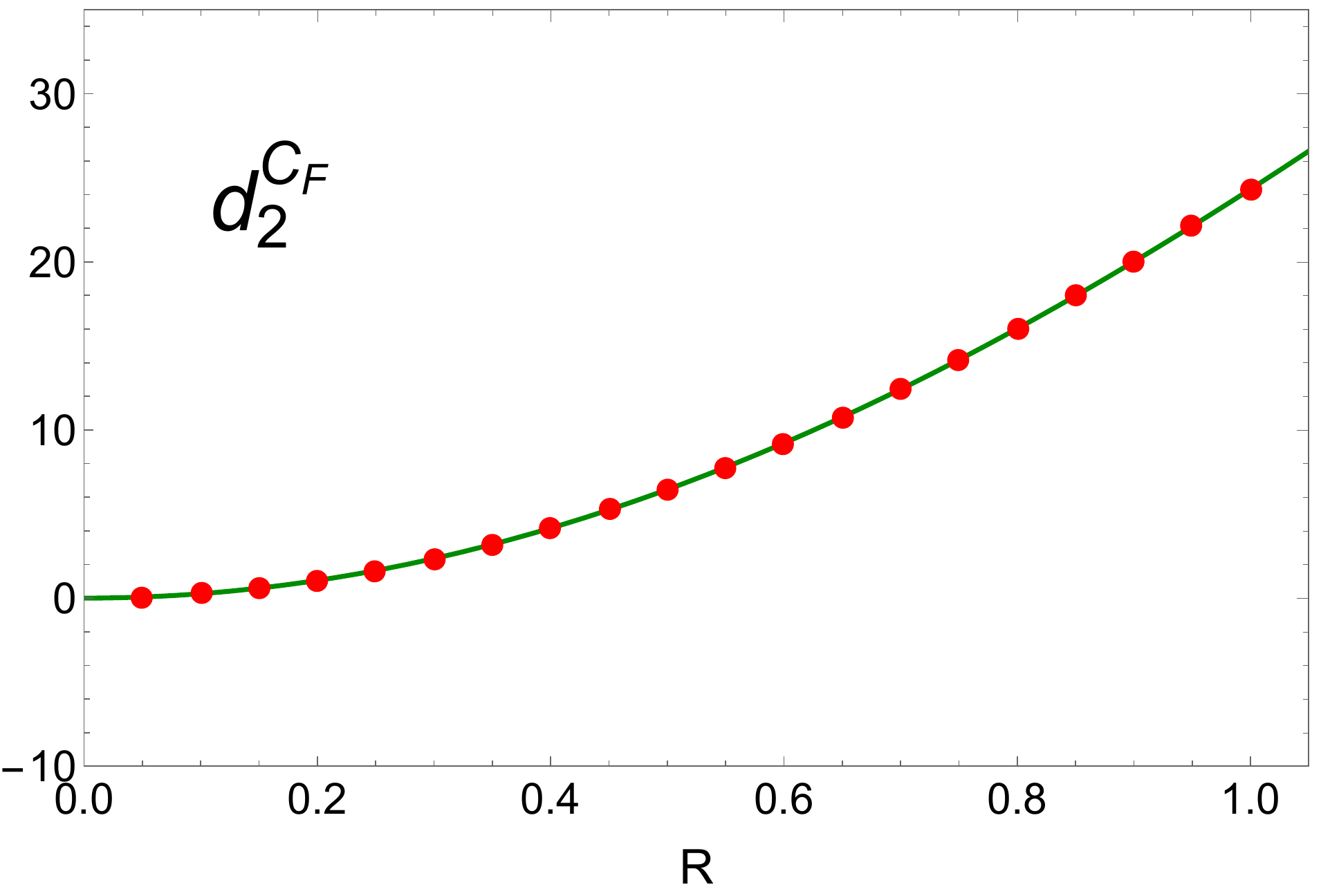}
\caption{Two-loop collinear anomaly exponent of the $p_T$-veto soft function. 
The (red) dots represent our numbers and the solid (green) 
lines are the fit functions from~\cite{Banfi:2012yh,Becher:2013xia,Stewart:2013faa}.}
\label{fig:pTveto}
\end{figure}

We finally consider the standard jet-veto soft function that is based on
transverse momenta. This is our first example of a SCET-2 observable with $n=0$,
for which we compute the two-loop anomaly exponent $d_2$ that was defined in 
(\ref{eq:d1d2}). The anomaly exponent has a similar colour decomposition,
\begin{align} 
d_2 &= d_2^{C_A} \;C_F C_A 
+ d_2^{n_f} \;C_F T_F n_f
+ d_2^{C_F} \;C_F^2\,,
\end{align}
and the $p_T$-veto soft function again renormalises multiplicatively in cumulant space.
At NLO one finds $d_1=0$ and at NNLO one obtains a similar picture as for the rapidity-dependent 
jet vetoes. Our results for $d_2$ are shown in Figure~\ref{fig:pTveto}
together with fitting functions that were obtained in previous 
calculations~\cite{Banfi:2012yh,Becher:2013xia,Stewart:2013faa}. Our
results are once more in agreement with 
the existing results.

\section{Conclusions}

We have developed an algorithm that allows for an automated calculation of arbitrary 
two-loop soft functions that are defined in terms of two back-to-back light-like 
Wilson lines. The algorithm has been implemented in the custom program {\tt SoftSERVE},
which can be used to determine both SCET-1 and SCET-2 soft functions. We illustrated 
the use of {\tt SoftSERVE} with a few examples that are sensitive to jet clustering effects
and hence violate the NAE theorem. Our results for the soft-drop jet-grooming soft 
function superseed previous determinations of the two-loop anomalous dimension that 
were valid only for specific values of the jet-grooming parameter. We plan to publish 
{\tt SoftSERVE} in the near future.

\acknowledgments

This work was supported in parts by the Deutsche Forschungsgemeinschaft (DFG) within 
Research Unit FOR 1873 (GB) and by the Swiss National Science Foundation (SNF) via 
grant CRSII2\_16081 (RR). JT acknowledges research and travel support from DESY.
RR and JT would like to thank the particle theory group in Siegen for its hospitality 
and support. GB would like to thank the organisers of RADCOR 2017 for creating a 
pleasant and stimulating workshop atmosphere.

\end{document}